\documentclass[review]{elsarticle}

\usepackage{lineno}
\modulolinenumbers[5]
\usepackage{balance}
\usepackage{comment}
\usepackage{multirow}
\usepackage{tikz,pgfplots,pgfplotstable,booktabs}
\usepackage{eurosym}
\usepackage[mathscr]{euscript}
\usepackage{amsmath,amssymb,amsfonts}
\usepackage{algorithmic}
\usepackage{graphicx}
\usepackage{textcomp}
\usepackage[utf8]{inputenc}
\usepackage{url}
\usepackage{bm}

\usepackage{tikz}
\usetikzlibrary{arrows,decorations.pathmorphing,backgrounds,fit,positioning,shapes.symbols,chains, shapes,snakes}

\newenvironment{ldescription}[1]
  {\begin{list}{}%
   {\renewcommand\makelabel[1]{##1\hfill}%
   \settowidth\labelwidth{\makelabel{#1}}%
   \setlength\leftmargin{\labelwidth}
   \addtolength\leftmargin{\labelsep}}}
  {\end{list}}

\journal{Journal of \LaTeX\ Templates}

\begin{document}

\begin{frontmatter}

\title{Forecasting the Price-Response of a Pool of Buildings via Homothetic Inverse Optimization}


\cortext[mycorrespondingauthor]{Corresponding author}

\author[a]{Ricardo Fernández-Blanco\corref{mycorrespondingauthor}}
\ead{Ricardo.FCarramolino@gmail.com}

\author[a,b]{Juan Miguel Morales}
\ead{Juan.Morales@uma.es}

\author[a,c]{Salvador Pineda}
\ead{spinedamorente@gmail.com}

\address[a]{OASYS research group, University of Malaga, Malaga, Spain}
\address[b]{Department of Applied Mathematics, University of Malaga, Malaga, Spain}
\address[c]{Department of Electrical Engineering, University of Malaga, Malaga, Spain}

\begin{abstract}
This paper focuses on the day-ahead forecasting of the aggregate power of a pool of smart buildings equipped with thermostatically-controlled loads. We first propose the modeling of the aggregate behavior of its power trajectory by using a geometric approach. Specifically, we assume that the aggregate power is a \textit{homothet} of a \textit{prototype} building, whose physical and technical parameters are chosen to be the \textit{mean} of those in the pool. This allows us to preserve the building thermal dynamics of the pool. We then apply inverse optimization to estimate the homothetic parameters with bilevel programming. The lower level characterizes the price-response of the ensemble by a set of marginal utility curves and a homothet of the \textit{prototype} building, which, in turn, are inferred in the upper-level problem. The upper level minimizes the mean absolute error over a training sample. This bilevel program is transformed into a regularized nonlinear problem that is initialized with the solution given by an efficient heuristic procedure. This heuristic consists in solving two linear programs and its solution is deemed a suitable \textit{proxy} for the original bilevel problem. The results have been compared to state-of-the-art methodologies.
\end{abstract}


\end{frontmatter}


\section*{Nomenclature}
The main notation used throughout the text is stated below for quick reference. Matrices are defined in bold and upper-case, vectors are indicated in bold and lower-case, superscript ${\cdot}^{\prime}$ means \textit{observed}, and symbol $\widehat{\cdot}$ refers to an \textit{estimated} parameter. Other symbols are defined as required. 

\subsection*{Sets and Indices}
\begin{ldescription}{$xxxxx$}
\item [$\mathcal{B}$] Set of blocks, indexed by $b = 1 \ldots n_B$.
\item [$\mathcal{D}$] Set of days, indexed by $d$.
\item [$\mathcal{H}$] Set of hours, indexed by $h = 1 \ldots n_H$.
\item [$\Omega^{p/a/i}$] Set of physical- and technical-related parameters for the prototype/aggregate/individual building, respectively.
\item [$\Phi^{p/a}$] Set of model-related parameters for the prototype/aggregate building, respectively. 
\item [$i$] Index for individual building.
\item [$r$] Index for regressor.
\end{ldescription}

\subsection*{Parameters}
\begin{ldescription}{$xxxxx$}
\item [$a_1$] Energy dissipation rate. 
\item [$a_2$] Parameter defining the product of $\eta \cdot R$. 
\item [$C$] Thermal capacitance.
\item [$n_B$] Number of power blocks.
\item [$n_H$] Number of hours in a day.
\item [$P$] Rated power of a thermostatically-controlled load.
\item [$R$] Thermal resistance. 
\item [$UA$] Heat transfer coefficient between the room air and the ambient.
\item [$\iota$] Parameter used in the regularization technique.
\item [$\eta$] Coefficient of performance of a thermostatically-controlled load. 
\item [$\theta_{0, d}$] Initial indoor air temperature in day $d$.
\item [$\theta^r$] User-specified temperature set-point.
\item [$\hbar$] Heterogeneity factor.
\item [$\delta$] Discretization period.
\item [$\Delta$] Half of the temperature deadband.
\item [$\boldsymbol{c^s}$] Vector of penalty costs for slack variables, where the $h$th component is $c^s_h$.
\item [$\boldsymbol{\lambda}_{d}$] Electricity price in day $d$, where the $h$th component is $\lambda_{h, d}$.
\item [$\boldsymbol{\theta}_{d}^{\boldsymbol{amb}}$] Outdoor air temperature in day $d$, where the $h$th component is $\theta^{amb}_{h, d}$.
\item [$\boldsymbol{A}, \boldsymbol{B}$]  Matrices associated with the matricial form of the building's discrete dynamics. 
\item [$\boldsymbol{Z}_d$] Matrix of regressors in day $d$, where each component is expressed as $z_{h,r,d}$ at hour $h$ and for regressor $r$.
\item [$\boldsymbol{\Lambda}$] The inverse of matrix $\boldsymbol{A}$.
\end{ldescription}

\subsection*{Other symbols}
We remark that most of these symbols, except for the acronyms MAE and RMSE, take on the role of parameters in the forecasting model, while acting as variables in the (bilevel) inverse optimization problem.

\begin{ldescription}{$xxxxx$}
\item [$\beta$] Scaling factor of the homothetic transformation.
\item [$\boldsymbol{\nu}_b$] Intercept for the marginal utility of block $b$, where component $h$th is $\nu_{b,h}$.
\item [$\text{MAE}$] Mean absolute error.
\item [$\text{RMSE}$] Root mean square error.
\item [$\boldsymbol{c}^{p/a}_d$] Vector representation of the building's initial conditions in day $d$ for the prototype/aggregate building, where component $h$th is $c^{p/a}_{h, d}$.
\item [$\overline{\boldsymbol{e}}_{b,d}$] Length of the power block $b$ in day $d$, where component $h$th is $\overline{e}_{h,b,d}$.
\item [$\boldsymbol{p}^{p/a}_{d}$] Power of the thermostatically-controlled load in day $d$ for the prototype/aggregate building, where component $h$th is $p^{p/a}_{h, d}$.
\item [$\boldsymbol{m}_{b,d}$] Marginal utility of block $b$ and day $d$, where component $h$th is $m_{b,h,d}$. 
\item [$\boldsymbol{p}^{a}_{b,d}$] Power of the thermostatically-controlled load in block $b$ and day $d$ for the aggregate building, where component $h$th is $p^{a}_{h,b,d}$.
\item [$\underline{\boldsymbol{p}}^{p/a}_{d}$] Minimum power of the thermostatically-controlled load in day $d$ for the prototype/aggregate building, where component $h$th is $\underline{p}^{p/a}_{h, d}$.
\item [$\overline{\boldsymbol{p}}^{p/a}_{d}$] Maximum power of the thermostatically-controlled load in day $d$ for the prototype/aggregate building, where component $h$th is $\overline{p}^{p/a}_{h, d}$.
\item [$\boldsymbol{s}^a_{d}$] Slack variable for the temperature-related constraints of the aggregate building in day $d$, where component $h$th is $s^a_{h, d}$.
\item [$\boldsymbol{t}^{p/a}_d$] Vector representation of the component of the building's discrete dynamics associated with the ambient temperature in day $d$ for the prototype/aggregate building, where component $h$th is $t^{p/a}_{h, d}$.
\item [$\boldsymbol{\theta}^{p/a}_{d}$] Indoor air temperature in day $d$ for the prototype/aggregate building, where component $h$th is $\theta^{p/a}_{h, d}$.
\item [$\underline{\boldsymbol{\theta}}^{p/a}_{d}$] Minimum indoor air temperature in day $d$ for the prototype/aggregate building, where component $h$th is $\underline{\theta}^{p/a}_{h, d}$.
\item [$\overline{\boldsymbol{\theta}}^{p/a}_{d}$] Maximum indoor air temperature in day $d$ for the prototype/aggregate building, where component $h$th is $\overline{\theta}^{p/a}_{h, d}$.
\item [$\boldsymbol{\tau}$] Translation factor of the homothetic transformation, where component $h$th is $\tau_h$.
\item [$\boldsymbol{\rho}$] Vector of coefficients relative to the affine dependence of marginal utility on regressors, where component $r$th is $\rho_r$.
\end{ldescription}

\section{Introduction}
\label{sec:introduction}
Distributed energy resources (DERs), such as distributed generators, electric vehicles, energy batteries or demand response programs, are constantly growing every year and play a crucial role in the provision of multiple benefits to the power system \cite{pinson2014benefits}. In this paper, we focus on a recently popular DER, namely, an ensemble of smart buildings \cite{junker2018characterizing}. This pool of buildings may efficiently utilize their thermal capacity while keeping the indoor air temperature at user-defined comfort levels in order to provide some degree of flexibility to the power system, by shifting their load in time or reducing the peak demand. In addition, this flexibility may allow its participation in a day-ahead electricity market or could even be viewed as a non-wire alternative to capacity expansion. However, as with any load in the electricity system, its prediction is key to fully exploit the benefits that can bring to the power system operation and planning \cite{shahidehpour2003market}. 

Recently, machine learning has become a popular forecasting technique in many scientific fields such as agriculture \cite{liakos2018machine}, medicine \cite{shkolyar2019augmented}, and even power systems \cite{ibrahim2020machine}. Ibrahim \emph{et al.} \cite{ibrahim2020machine} give a thorough overview of machine learning techniques applied for electricity load and price prediction, renewable generation forecasting, fault detection, failure analysis, among other applications for smart grids. Lago \emph{et al.} \cite{lago2018forecasting} compare the performance of various black-box models including traditional statistical techniques and four deep learning approaches to predict day-ahead electricity prices. This study demonstrates that machine learning techniques outperform commonly used statistical models for forecasting day-ahead electricity prices. However, as pointed out in \cite{lago2018forecasting}, machine learning techniques require the tuning of model-specific hyperparameters (e.g. number of layers, dropout, activation function, among others). 

This paper is focused on price-responsive load forecasting, which has been also studied in the technical literature by using a plethora of black-box models \cite{corradi2012controlling, jeong2020short, yun2008rbf}. For instance, authors in \cite{corradi2012controlling} propose the use of statistical models such as auto-regressive models with exogenous inputs (also known as ARX) to forecast the dynamics of the price-elasticity of price-responsive consumers. They demonstrate the applicability of the model by using data from price-responsive heating systems from the Olympic Peninsula project \cite{hammerstrom2008pacific}. Authors in \cite{jeong2020short} apply a logistic mixture vector auto-regressive model to forecast daily electric curves of buildings with a pattern history. According to their results, the proposed statistical model outperforms other machine learning techniques (e.g. support vector machines) for the analyzed case studies. Reference \cite{yun2008rbf} demonstrates how neural networks and fuzzy systems can be combined to predict real-time electrical loads. However, black-box models are purely data-driven and, as a consequence, the physics of the load to be predicted are ignored. In other words, this kind of models would neglect the technical and physical constraints governing DERs, e.g. thermostatically-controlled loads or electric vehicles. Hence, black-box models lack interpretability and explainability \cite{pintelas2020grey, arendt2018comparative}. For this reason, in this work, we advocate for grey-box models, which typically rely on a reduced physical model that is estimated or complemented with data, this way featuring a balance between interpretability and scalability.

Grey-box models have been put forward in \cite{lu2015modeling} and \cite{Saez-Gallego2018} to predict the price-response of a pool of buildings . Reference \cite{lu2015modeling} takes into account the physical model of buildings to forecast their energy consumption, however its application is limited to predict individual buildings' loads instead of forecasting an aggregate load of a pool of buildings. Recently, \cite{Saez-Gallego2018} proposes a novel inverse optimization (IO) approach to statistically forecast the aggregate load of a pool of price-responsive buildings in an hour-ahead setting. In that paper, the authors characterize the response of the load to the price by means of an optimization problem. The limitations of the model proposed in \cite{Saez-Gallego2018} are threefold: (i) the methodology is based on heuristics, (ii) the optimization models are tailored to single-step forecasts, and thus its use for multi-step forecasting is inappropriate, and, as a consequence, (iii) the building thermal dynamics are disregarded in the forecasting process. In this paper, we also apply IO to forecast the aggregate response to the electricity price of a pool of buildings. However, unlike \cite{Saez-Gallego2018}, our goal is to predict it in a \emph{day-ahead} framework while also incorporating the building thermal dynamics into the optimization process. To do that, we put forward a novel IO approach that makes use of an homothetic transformation to conveniently reduce the complexity of the target forecasting model.

The goal of an IO problem is to infer the optimization model parameters given a set of observed decision variables or measurements collected by an observer. Recent advances on IO can be found in \cite{aswani2018, esfahani2018data, bertsimas2015data,chan2020inverse,ghobadi2020inferring}, and references therein. Aswani \emph{et al.} \cite{aswani2018} devise a statistically consistent IO methodology based on the transformation of a bilevel problem into a single-level equivalent when data is corrupted by noise. The authors show the performance of the proposed IO methodology to estimate cost vector parameters compared to existing heuristics with synthetic and empirical data. In a more general mathematical context, Esfahani \emph{et al.} \cite{esfahani2018data} prescribe a data-driven approach based on distributionally robust IO to tackle observed measurements with imperfect information. From a different angle, Bertsimas \emph{et al.} \cite{bertsimas2015data} apply IO to equilibrium models by using an inverse variational inequality formulation and demonstrate its predictive performance in two applications related to demand and congestion function estimation. Recently, authors in \cite{chan2020inverse} estimate the feasible region of linear and robust linear optimization problems by using IO with noise-free data, whereas authors in \cite{ghobadi2020inferring} impute a linear feasible region with a general IO methodology so that some noise-free observations become feasible and others become optimal provided that the cost vector is known.

IO is used in various applications in the technical literature \cite{zhou2011designing,ruiz2013,risanger2020inverse,saez2016data,lu2018data,contreras2018tractable,kovacs2020inverse}. The authors in \cite{zhou2011designing} apply IO in the context of generation expansion planning to find an effective incentive policy and those in \cite{ruiz2013} estimate rival marginal offer prices for a strategic producer in a network-constrained day-ahead market by using IO. The benefits and barriers of an inverse equilibrium problem are discussed in \cite{risanger2020inverse} for two applications of a Nash-Cournot game between power producers in electricity markets. In this inverse equilibrium problem, the aim is to estimate all parameters related to the original equilibrium model.    

Within the context of smart buildings, IO has also been applied to characterize price-responsive consumers in \cite{saez2016data,lu2018data}. More specifically, bilevel programming is used in \cite{saez2016data} to construct an IO framework whereby the market bid parameters of a pool of price-responsive households (e.g., step-wise marginal utility functions, maximum load pick-up and drop-off rates, and maximum and minimum power
consumption bounds) are inferred. Although \cite{saez2016data} accounts for a refined model of the aggregate load of the households by including the ramping rates, it still neglects the inertia governing the households' thermal consumption. IO is also applied in \cite{lu2018data} to estimate the demand response characteristics of price-responsive consumers, as similarly done in \cite{saez2016data}, and thus sharing the same shortcoming. The authors in \cite{contreras2018tractable} describe a data-driven method to empirically estimate a robust feasible region of a pool of buildings. Authors in \cite{kovacs2020inverse} infer the parameters of electricity consumer models with multiple deferrable loads from historic data by using IO. The proposed IO model is reformulated to a quadratically constrained quadratic program. The resulting model is solved by means of successive linear programming. Although the results are promising due to the prediction performance for single loads, the convergence properties of the successive linear programming approach are arguably, according to the authors. However, the thermal dynamics of the buildings are once again ignored from the estimation procedure in these works \cite{saez2016data,lu2018data,contreras2018tractable,kovacs2020inverse}.

One of the main contributions of this paper is the application of a geometric approach, i.e. we resort to the concept of \textit{homothety},  to characterize the price-response of the ensemble of buildings for forecasting purposes. A homothety is a spatial transformation of an affine space. Hence, we assume that the feasible region of a pool of buildings can be represented as a homothet of a chosen \textit{prototype} building by means of a dilation factor and a translation vector, namely the homothetic parameters. The homothetic representation of an aggregate of buildings has been first proposed in \cite{zhao2017geometric}. Specifically, they put forward the modeling of the aggregate flexibility of a pool of thermostatically-controlled loads by using a geometric approach. The thrust of that paper was to derive sufficient and necessary virtual battery models that can be approximated by homothets of a virtual battery prototype. The authors demonstrated the benefits of such homothetic representation in the provision of flexibility for regulation services. To the best of our knowledge, this is the first time that a homothetic representation of an aggregate load has been applied for forecasting purposes. Consequently, we only rely on the estimation of the homothetic parameters to shape the aggregate feasible region of the pool, thus considerably reducing the computational complexity of the estimation algorithm and avoiding an undesirable overfitting. This work contributes to the technical literature as follows: 

\begin{itemize}
    \item From a modeling perspective, we propose a novel day-ahead forecasting technique for a pool of buildings via homothetic inverse optimization. The aggregate price-response is characterized by a set of marginal utility curves and a homothet of a \textit{prototype} building. As novel distinctive features compared to the work in \cite{Saez-Gallego2018}, this geometric approach endogenously accounts for the aggregate building thermal dynamics and allows us to solely rely on the estimation of two homothetic parameters and a set of marginal utility curves. We then apply IO to infer them through a given forecasting problem\footnote{Also known as forward problem in the jargon of inverse optimization.} \textit{mimicking} the price-response of the pool. Our approach, therefore, drastically reduces the complexity of the price-response model to be statistically estimated, while still capturing the thermal dynamics of the ensemble of buildings, unlike the works in \cite{Saez-Gallego2018,saez2016data,lu2018data,contreras2018tractable,kovacs2020inverse}.
    
    \item The application of IO gives rise to a bilevel programming problem. We then propose the transformation of this bilevel program into a regularized nonlinear model which can be readily solved by nonlinear commercial solvers. To avoid meaningless local optimal solutions, we initialize this regularized model with the solution given by an efficient heuristic procedure. 
    \item The proposed forecasting technique has been compared with existing methodologies emphasizing its benefits for different degrees of heterogeneity among buildings.
\end{itemize}

The contribution of this paper is focused on the forecasting of the aggregate load of an ensemble of buildings by using a geometric approach, which is quite interpretable, in a day-ahead setting. Apart from the forecast of the aggregate price-response, the proposed approach allows for the derivation of a bidding curve that can be used in day-ahead electricity markets. Besides, the geometric interpretation of the ensemble of smart buildings by means of a prototype building may be used for accurately representing their power trajectory in operation problems at distribution level. Obviously, some deviations could be observed in intra-day applications. However, the proposed approach may be easily adapted for shorter time periods by adequately adjusting the length of the optimization horizon.

The paper is outlined as follows. 
Section \ref{sec:derivation_fwm} provides the derivation of the feasible region for a pool of buildings by using a homothet of a building prototype. Section \ref{sec:inverse_opt_methodology} presents the proposed IO methodology based on a bilevel program. Section \ref{sec:comparison_methodologies} describes the forecasting methodologies used to benchmark our proposal. Section \ref{sec:case_study} provides insightful results. Conclusions are duly drawn in Section \ref{sec:conclusion}. Finally, \ref{sec:two_step_estimation} provides the mathematical formulations for the proposed two-step heuristic estimation procedure used for initialization of the proposed single-level nonlinear program.

\section{Derivation of the Forecasting Model}
\label{sec:derivation_fwm}
In Section \ref{sec:building_prototype}, we first present the feasible region of a prototype building which can be representative of the ensemble of buildings. Subsequently, in Section \ref{sec:aggregate_building_model}, we provide the feasible region of an aggregate building which is built upon the prototype building by using the concept of \emph{homothet}. Finally, in Section \ref{sec:forecasting_model}, we derive the forecasting model. 

\subsection{Building Prototype}
\label{sec:building_prototype}
We consider that the prototype building is the one representing the \textit{average} behavior of those in the pool. To do that, we model the prototype building as a single thermostatically-controlled load characterized by a thermal resistance, $R = 1/UA$, being $UA$ the heat transfer coefficient between the room air and the ambient, and the thermal capacitance of the room air, $C$. In addition, we assume that the building is equipped with a cooling system with a rated power $P$ and a coefficient of performance $\eta$. Bearing in mind both the temperature comfort bounds by the building's occupants and the technical power limits of the cooling device, the feasible region of the prototype building for $n_H$ time periods within a day can be mathematically expressed as:
\begin{subequations}
\label{feasible_region_prototype}
\begin{align}
& \theta_h^p = a_1 \theta_{h-1}^p + (1-a_1)\left[ \theta^{amb}_{h} - a_2 p_{h}^p\right], \quad  \forall h \in \mathcal{H} \label{fr_one_1}  \\
& \underline{\theta}_h^p \leq \theta_h^p \leq \overline{\theta}_h^p, \quad \forall h \in \mathcal{H} \label{fr_one_2}  \\
& \underline{p}_h^p \leq p_{h}^p \leq \overline{p}_h^p, \quad \forall h \in \mathcal{H}, \label{fr_one_3} 
\end{align}
\end{subequations}
\noindent where $\underline{\theta}_h^p = \theta^{r} - \Delta$ and $\overline{\theta}_h^p = \theta^{r} + \Delta$, being $\theta^{r}$ the user-specified temperature set-point and $\Delta$ the half of the temperature deadband; and $a_1 = 1 - \dfrac{\delta}{RC}$ and $a_2 = \eta R$, with $\delta$ being the discretization period. Besides, $\underline{p}_h^p = 0$ and $\overline{p}_h^p = P$. To sum up, the set of physical and technical parameters of the prototype building is $\Omega^p = \{R, C, \theta^{r}, \Delta,  \eta, \theta_0, P\}$. These values are assumed to be the average of the values for the same set of parameters corresponding to each building of the ensemble. In this section, we have dropped index $d$ for the sake of clarity. Note that, in equation \eqref{fr_one_1} and throughout the paper, we assume a linear system for modeling the building thermal dynamics of thermostatically-controlled loads as opposed to a nonlinear switching model. This assumption allows to simplify the analysis of the aggregate behavior of an ensemble of buildings and, when the population of buildings is large, the aggregate power demand of the nonlinear switching models can be accurately approximated by a linear system model \cite{zhao2017geometric}. 

Conveniently and following the notation in \cite{zhao2017geometric}, we can recast the thermal model \eqref{feasible_region_prototype} in matricial form: 
\begin{subequations}
\label{feasible_region_prototype_matricial}
\begin{align}
& \underline{\boldsymbol{p}}^p \leq \boldsymbol{p}^p \leq \overline{\boldsymbol{p}}^p \label{vb_1} \\
& \underline{\boldsymbol{\theta}}^p \leq \boldsymbol{\Lambda}\boldsymbol{B}\boldsymbol{p}^p + \boldsymbol{\Lambda}\boldsymbol{c}^p + \boldsymbol{\Lambda}\boldsymbol{t}^p \leq \overline{\boldsymbol{\theta}}^p , \label{vb_2} 
\end{align}
\end{subequations}

\noindent where $\boldsymbol{\Lambda}$ is the inverse of $\boldsymbol{A}$; $\boldsymbol{A} = \boldsymbol{I}_{n_H} + diag(-a_1; -1)$, wherein $\boldsymbol{I}_{n_H}$ is the identity matrix of dimension $n_H$ and $diag(-a_1; -1)$ is a matrix of dimension $n_H$ with values $-a_1$ on the lower subdiagonal; $\boldsymbol{B} = - a_2 \left( 1-a_1\right) \boldsymbol{I}_n$; $\boldsymbol{c}^p$ is the vector of initial conditions $[a_1\theta_0, 0, ..., 0]^T$, being superscript $T$ the transpose operator; and $\boldsymbol{t}^p$ is the vector related to the ambient temperature, i.e. $\boldsymbol{\theta^{amb}}\left( 1-a_1\right)$. We denote the set of model-related parameters for the building prototype as $\Phi^p = \{\boldsymbol{c}^p, \underline{\boldsymbol{p}}^p, \overline{\boldsymbol{p}}^p, \boldsymbol{t}^p, \underline{\boldsymbol{\theta}}^p, \overline{\boldsymbol{\theta}}^p\}$.

\subsection{Aggregate Building Model}
\label{sec:aggregate_building_model}
We can approximate the feasible region of the aggregation of buildings, for each day $d$, as another thermal building model, that is,
\begin{subequations}
\label{aggregate_building_model}
\begin{align}
& \underline{\boldsymbol{p}}^a_d \leq \boldsymbol{p}^a_d \leq \overline{\boldsymbol{p}}^a_d \label{const1_hvb}  \\
& \underline{\boldsymbol{\theta}}^a_d \leq \boldsymbol{\Lambda} \boldsymbol{B} \boldsymbol{p}^a_d + \boldsymbol{\Lambda} \boldsymbol{c}^a_d + \boldsymbol{\Lambda}\boldsymbol{t}^a_d \leq \overline{\boldsymbol{\theta}}^a_d. \label{const2_hvb} 
\end{align}
\end{subequations}

However, the set of model-related parameters of the pool of buildings associated with \eqref{aggregate_building_model}, i.e, $\Phi^a = \{\boldsymbol{c}^a_d, \underline{\boldsymbol{p}}^a_d, \overline{\boldsymbol{p}}^a_d, \boldsymbol{t}^a_d, \underline{\boldsymbol{\theta}}^a_d, \overline{\boldsymbol{\theta}}^a_d\}$, are unknown. One possibility would be to infer all these parameters from observations of the aggregate power of the pool of buildings. However, this is most likely to be a lost cause (due to unobservability issues), lead to overfitting, and  result in instability of the estimation algorithm. To overcome such difficulty, we assume that the aggregate feasible region is a \textit{homothet} of the prototype building, i.e., the power trajectory of the aggregate of buildings for each day $d$ can be expressed in terms of the power trajectory of the prototype building as follows: 
\begin{align}
\boldsymbol{p}^a_d = \beta \boldsymbol{p}^p_d + \boldsymbol{\tau}, \label{definition_homothet}
\end{align}

\noindent where $\beta > 0$ is a scaling factor, and $\boldsymbol{\tau}$ is a vector of translation factors. Expression \eqref{definition_homothet} is the formal definition of a homothet for the aggregate power in $\mathbb{R}^{n_H}$, i.e. vectors $\boldsymbol{p}^a_d$, $\boldsymbol{p}^p_d$, and $\boldsymbol{\tau}$ contains $n_H$ components. 

By using the definition of homothet \eqref{definition_homothet} and the prototype feasible region \eqref{feasible_region_prototype_matricial}, we can recast the feasible region of the aggregation in terms of the homothetic parameters $\beta$ and $\boldsymbol{\tau}$: 
\begin{subequations}
\label{homothetic_feasible_region}
\begin{align}
& \beta \underline{\boldsymbol{p}}^p_d + \boldsymbol{\tau}  \leq \boldsymbol{p}^a_d \leq \beta \overline{\boldsymbol{p}}^p_d + \boldsymbol{\tau} \label{const1_hvb2}  \\
& \beta \underline{\boldsymbol{\theta}}^p_d  + \boldsymbol{\Lambda} \boldsymbol{B}  \boldsymbol{\tau} \leq \boldsymbol{\Lambda} \boldsymbol{B} \boldsymbol{p}^a_d + \boldsymbol{\Lambda} \beta \left(\boldsymbol{c}^p_d \hspace{-0.1cm} + \hspace{-0.1cm} \boldsymbol{t}^p_d\right) \leq  \beta \overline{\boldsymbol{\theta}}^p_d + \boldsymbol{\Lambda} \boldsymbol{B} \boldsymbol{\tau}. \label{const2_hvb2} 
\end{align}
\end{subequations}

The feasible region of the homothetic aggregate building \eqref{homothetic_feasible_region} depends entirely on the homothetic parameters and the set of model-related parameters of the prototype building $\Phi^p$ (which are given), thus dramatically reducing the complexity of the model to be estimated and avoiding the undesirable overfitting effect. To the best of the authors' knowledge, this is the first time in the literature that such a geometric approach is used to drastically simplify the task of forecasting the price-responsive aggregate power of a pool of buildings via IO, as explained below.

\subsection{Forecasting Model}
\label{sec:forecasting_model}
Let us assume that the feasible region of the ensemble of buildings is a homothetic representation of a prototype building and that the utility function of the pool is a step-wise price function with $n_B$ blocks. Under these assumptions and given the electricity prices, the forecasting model for each day $d$ can be mathematically expressed as: 
\begin{subequations}
\label{forecasting_model}
\begin{align}
&\max_{\boldsymbol{p}^a_{b, d}, \boldsymbol{s}^a_d} \quad \sum_{b \in \mathcal{B}}\bigl(\boldsymbol{m}_{b, d}^T - \boldsymbol{\lambda}^T_d \bigr) \boldsymbol{p}^{a}_{b, d}  - \boldsymbol{c}^{s, T} \boldsymbol{s}^{a}_d   \label{fo_fwp} \\
& \text{subject to:} \notag\\
& \beta \underline{\boldsymbol{p}}^p_d + \boldsymbol{\tau} \leq \sum_{b \in \mathcal{B}} \boldsymbol{p}^a_{b, d} \leq \beta\overline{\boldsymbol{p}}^p_d + \boldsymbol{\tau} : (\underline{\boldsymbol{\epsilon}}_d, \overline{\boldsymbol{\epsilon}}_d) \label{const1_fwp}  \\
&  \beta \underline{\boldsymbol{\theta}}^p_d  + \boldsymbol{\Lambda} \boldsymbol{B}  \boldsymbol{\tau} - \boldsymbol{s}^a_d \leq \sum_{b \in \mathcal{B}}  \boldsymbol{\Lambda} \boldsymbol{B} \boldsymbol{p}^a_{b, d} \hspace{-0.1cm} + \hspace{-0.1cm} \boldsymbol{\Lambda} \beta \left(\boldsymbol{c}^p_d \hspace{-0.1cm} + \hspace{-0.1cm} \boldsymbol{t}^p_d\right) : (\underline{\boldsymbol{\kappa}}_d) \label{const2_fwp}  \\
& \sum_{b \in \mathcal{B}} \boldsymbol{\Lambda} \boldsymbol{B} \boldsymbol{p}^a_{b, d} \hspace{-0.1cm} + \hspace{-0.1cm} \boldsymbol{\Lambda} \beta \left(\boldsymbol{c}^p_d \hspace{-0.1cm} + \hspace{-0.1cm} \boldsymbol{t}^p_d\right) \leq \beta \overline{\boldsymbol{\theta}}^p_d + \boldsymbol{\Lambda} \boldsymbol{B} \boldsymbol{\tau} +  \boldsymbol{s}^a_d : (\overline{\boldsymbol{\kappa}}_d) \label{const3_fwp}  \\
&  0 \leq \boldsymbol{p}^a_{b, d} \leq \overline{\boldsymbol{e}}_{b, d}: (\underline{\boldsymbol{\phi}}_{b, d}, \overline{\boldsymbol{\phi}}_{b, d}), \quad \forall b \in \mathcal{B} \label{const4_fwp}\\
& \boldsymbol{s}^a_d \geq 0: (\boldsymbol{\varphi}_d),  \label{const5_fwp}
\end{align}
\end{subequations}

\noindent where $\boldsymbol{m}_{b, d}$, $\boldsymbol{\lambda}_d$, and $\boldsymbol{c}^s$ are the vectors of marginal utilities, electricity prices, and penalty costs. The dual variables are shown in parentheses after a colon next to the corresponding constraints. 

The objective function \eqref{fo_fwp} aims to maximize the welfare of the pool of buildings while minimizing the slack variables associated with the evolution of the building thermal dynamics. Constraints \eqref{const1_fwp}--\eqref{const3_fwp} are almost identical to the homothetic representation of the aggregate feasible region \eqref{homothetic_feasible_region}. Without loss of generality, we have incorporated some degree of flexibility into the forecasting model by: (i) modeling step-wise marginal utility functions to adequately learn the price-response of the pool of buildings, and (ii) including the slack variable in the temperature-related constraints \eqref{const2_fwp}--\eqref{const3_fwp} to capture the infeasibilities that the (approximate) modeling of the building thermal dynamics may cause. Constraints \eqref{const4_fwp} impose lower and upper bounds on the aggregate power per block $b$, being $\overline{\boldsymbol{e}}_{b, d}$ the length of the power per block $b$ in day $d$. Finally, constraints \eqref{const5_fwp} set the non-negative character of slack variables. 

As previously mentioned, the feasible region of the ensemble is parameterized in terms of the homothetic parameters $\beta$ and $\boldsymbol{\tau}$. Therefore, in this problem, the vector of marginal utilities $\boldsymbol{m}_{b, d}$, as well as the homothetic parameters $\beta$ and $\boldsymbol{\tau}$ are parameters to be estimated through IO, as explained in Section \ref{sec:inverse_opt_methodology}. 

\section{Inverse Optimization Methodology}
\label{sec:inverse_opt_methodology}
In this section, we describe the proposed IO methodology to infer the parameters $\boldsymbol{m}_{b, d}$, $\beta$, and $\boldsymbol{\tau}$ of the forecasting model \eqref{forecasting_model}. This methodology is based on bilevel programming, which has been widely used in the technical literature to model hierarchical optimization problems \cite{deSouza2020optimal,cornelusse2019community,fernandez2015bilevel}.  First, we present our bilevel program and its transformation into a parametric (or regularized) nonlinear single-level equivalent program that can be solved by commercial solvers. 
Then, we thoroughly explain the steps of the proposed approach. 

\begin{figure}[h]
\centering
\begin{tikzpicture}
[node distance = 2.4cm, auto,font=\footnotesize,
every node/.style={node distance=2.8cm},
force/.style={rectangle, draw, fill=black!1, inner sep=12pt, text width=6cm, text badly centered, minimum height=1.5cm, minimum width=8.9cm, font=\bfseries\footnotesize\sffamily}]
\node [force] (UL) {Upper-level problem \eqref{fo_bilevel}--\eqref{const2_bilevel} \\  (Minimize the MAE of the aggregate power)};
\node [force, below of=UL] (LL) {Lower-level problems \eqref{const3_bilevel} for each day $d$ \\  (Forecasting model)};
\draw [->,thick] (-1,-0.80) -- (-1,-2.0) node [midway, left] {$\boldsymbol{m}_{b,d},\beta, \boldsymbol{\tau}$}  ;
\draw [->,thick] (1.7,-2.0) -- (1.7,-0.8) node [midway, right] {$\boldsymbol{p}_{b,d}^{a}$} ;
\end{tikzpicture}

\caption{A conceptual diagram of the interfaces of the bilevel problem.}
\label{fig:sketch}
\end{figure}
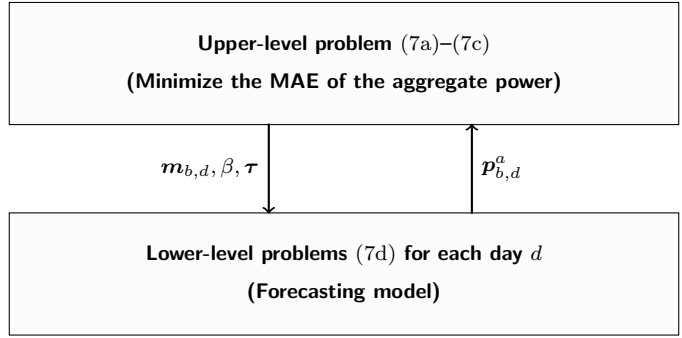

\subsection{Bilevel Problem}
\label{subsec:bilevel_problem}
The bilevel problem consists of two optimization levels, as depicted in Fig. \ref{fig:sketch}. In the upper-level problem, we seek to minimize the mean absolute error (MAE) of the aggregate power of the ensemble of buildings. This level provides the marginal utilities $\boldsymbol{m}_{b,d}$ as well as the homothetic parameters $\beta$ and $\boldsymbol{\tau}$ needed to build the homothetic representation in the lower-level problem. In contrast, in the lower level, we solve the maximization of the welfare of the pool of buildings and the minimization of the violations related to the building thermal dynamics for each day of the training set. In turn, the lower level passes the values of the optimal aggregate power on to the upper-level problem.

Let us denote the vector of observed aggregate power in day $d$ as $\boldsymbol{p}^{a^{\prime}}_{d}$. Therefore, the bilevel problem can be formulated as follows: 
\begin{subequations}
\label{bilevel_formulation}
\begin{align}
&\min_{\boldsymbol{m}_{b,d},\boldsymbol{p}_{b,d}^{a},\boldsymbol{s}_{d}^{a},\beta, \boldsymbol{\tau}, \boldsymbol{\nu}_b, \boldsymbol{\rho}} \quad \sum_{d \in \mathcal{D}} \Bigl|\Bigl| \sum_{b \in \mathcal{B}} \boldsymbol{p}_{b, d}^{a} - \boldsymbol{p}_{d}^{a^{\prime}}\Bigr|\Bigr|_1  \label{fo_bilevel} \\
& \text{subject to:} \notag\\
& \boldsymbol{m}_{b, d} = \boldsymbol{\nu}_b + \boldsymbol{Z}_{d}\boldsymbol{\rho} , \quad \forall b \in \mathcal{B}, d \in \mathcal{D} \label{const1_bilevel}  \\
& \boldsymbol{\nu}_b \geq \boldsymbol{\nu}_{b+1}, \quad \forall b < n_B \label{const2_bilevel}\\
& \text{Lower-Level Problem \eqref{forecasting_model},} \quad \forall d \in \mathcal{D}.  \label{const3_bilevel} 
\end{align}
\end{subequations}

On the one hand, the upper-level problem \eqref{fo_bilevel}-\eqref{const2_bilevel} minimizes the absolute error of the estimated aggregate power of the pool with respect to the observed one, as given by \eqref{fo_bilevel}. Constraints \eqref{const1_bilevel} impose linear regression functions, with $\boldsymbol{\nu}_b$ and $\boldsymbol{\rho}$ as the coefficients to be estimated, so that the marginal utilities are related to the regressors. We assume that vector $\boldsymbol{\nu}_b$ is time invariant, i.e. all components $\nu_{b, h}$ are identical. Constraints \eqref{const2_bilevel} set the marginal utilities to be monotonically non-increasing, as commonly done in electricity markets \cite{omie}. The lower-level problems \eqref{const3_bilevel} are essentially the forecasting problem \eqref{forecasting_model} for each day $d$. These lower levels are solely parameterized in terms of the marginal utilities $\boldsymbol{m}_{b, d}$, as well as the homothetic parameters $\beta$ and $\boldsymbol{\tau}$, and thus rendering the lower levels as linear programs. 
Therefore, we can apply the Karush-Khun-Tucker necessary optimality conditions to the lower level and apply the regularization described in \cite{scholtes2001convergence, pineda2018efficiently} to transform the original bilevel model \eqref{bilevel_formulation} into the following nonlinear single-level equivalent:
\begin{subequations}
\label{single_level_equivalent}
\begin{align}
&\min_{\Xi^{NRP}} \quad \sum_{d \in \mathcal{D}} \Bigl|\Bigl| \sum_{b \in \mathcal{B}} \boldsymbol{p}_{b, d}^{a} - \boldsymbol{p}_{d}^{a^{\prime}}\Bigr|\Bigr|_1  \label{fo_sle}\\
& \text{subject to:} \notag\\
& \boldsymbol{m}_{b, d} = \boldsymbol{\nu}_b + \boldsymbol{Z}_{d}\boldsymbol{\rho} , \quad \forall b \in \mathcal{B}, d \in \mathcal{D} \label{const1_sle} \\
& \boldsymbol{\nu}_b \geq \boldsymbol{\nu}_{b+1}, \quad \forall b < n_B \label{const1bis_sle}\\
& \beta \underline{\boldsymbol{p}}^p_d + \boldsymbol{\tau} \leq \sum_{b \in \mathcal{B}} \boldsymbol{p}^a_{b, d} \leq \beta\overline{\boldsymbol{p}}^p_d + \boldsymbol{\tau}, \quad \forall d \in \mathcal{D} \label{const2_sle}  \\
&  \beta \underline{\boldsymbol{\theta}}^p_d  + \boldsymbol{\Lambda} \boldsymbol{B}  \boldsymbol{\tau} - \boldsymbol{s}^a_d \leq \sum_{b \in \mathcal{B}}  \boldsymbol{\Lambda} \boldsymbol{B} \boldsymbol{p}^a_{b, d} \hspace{-0.1cm} + \hspace{-0.1cm} \boldsymbol{\Lambda} \beta \left(\boldsymbol{c}^p_d \hspace{-0.1cm} + \hspace{-0.1cm} \boldsymbol{t}^p_d\right), \quad \forall d \in \mathcal{D}  \label{const3_sle}  \\
& \sum_{b \in \mathcal{B}} \boldsymbol{\Lambda} \boldsymbol{B} \boldsymbol{p}^a_{b, d} \hspace{-0.1cm} + \hspace{-0.1cm} \boldsymbol{\Lambda} \beta \left(\boldsymbol{c}^p_d \hspace{-0.1cm} + \hspace{-0.1cm} \boldsymbol{t}^p_d\right) \leq \beta \overline{\boldsymbol{\theta}}^p_d + \boldsymbol{\Lambda} \boldsymbol{B} \boldsymbol{\tau} +  \boldsymbol{s}^a_d, \quad \forall d \in \mathcal{D} \label{const4_sle}  \\
& 0 \leq \boldsymbol{p}^a_{b,d} \leq \overline{\boldsymbol{e}}_{b,d}, \quad \forall b \in \mathcal{B}, d \in \mathcal{D} \label{const5_sle}\\
& \boldsymbol{s}^a_d \geq 0, \quad \forall d \in \mathcal{D} \label{const6_sle}\\
&-\underline{\boldsymbol{\epsilon}}_d + \overline{\boldsymbol{\epsilon}}_d - \boldsymbol{B}^T_d \boldsymbol{\Lambda}^T_d \underline{\boldsymbol{\kappa}}_d +\boldsymbol{B}^T_d \boldsymbol{\Lambda}^T_d \overline{\boldsymbol{\kappa}}_d  - \underline{\boldsymbol{\phi}}_{b,d} + \overline{\boldsymbol{\phi}}_{b,d} \notag \\
&=  \boldsymbol{m}_{b,d} - \boldsymbol{\lambda}_d, \quad \forall b \in \mathcal{B}, d \in \mathcal{D} \label{const7_sle}  \\
& \underline{\boldsymbol{\kappa}}_d +  \overline{\boldsymbol{\kappa}}_d + \boldsymbol{\varphi}_d= \boldsymbol{c}^s, \quad \forall d \in \mathcal{D} \label{const8_sle}  \\
& \underline{\boldsymbol{\epsilon}}_d, \overline{\boldsymbol{\epsilon}}_d, \underline{\boldsymbol{\kappa}}_d, \overline{\boldsymbol{\kappa}}_d,  \boldsymbol{\varphi}_{d} \geq 0, \quad \forall d \in \mathcal{D} \label{const9_sle}\\
& \underline{\boldsymbol{\phi}}_{b,d}, \overline{\boldsymbol{\phi}}_{b,d} \geq 0, \quad \forall b \in \mathcal{B}, d \in \mathcal{D} \label{const10_sle}
\end{align}
\begin{align}
& \sum_{d \in \mathcal{D}} \underline{\boldsymbol{\epsilon}}_d^T \Bigl( \sum_{b \in \mathcal{B}}\boldsymbol{p}^a_{b,d} - \beta \underline{\boldsymbol{p}}^p_d - \boldsymbol{\tau}  \Bigr)  +  \sum_{d \in \mathcal{D}} \overline{\boldsymbol{\epsilon}}_d^T \Bigl( \beta \overline{\boldsymbol{p}}^p_d + \boldsymbol{\tau} - \sum_{b \in \mathcal{B}}\boldsymbol{p}^a_{b,d}  \Bigr)   \notag\\
&+ \sum_{d \in \mathcal{D}} \underline{\boldsymbol{\kappa}}_d^T \Bigl(  \sum_{b \in \mathcal{B}} \boldsymbol{\Lambda} \boldsymbol{B} \boldsymbol{p}^a_{b,d} + \boldsymbol{\Lambda}_d \beta \boldsymbol{c}^p_d + \boldsymbol{\Lambda}_d \beta \boldsymbol{t}^p_d  -  \beta \underline{\boldsymbol{\theta}}^p_d  - \boldsymbol{\Lambda} \boldsymbol{B}  \boldsymbol{\tau} + \boldsymbol{s}_d^a  \Bigr)    \notag\\
&+ \sum_{d \in \mathcal{D}} \overline{\boldsymbol{\kappa}}_d^T \Bigl( \beta \overline{\boldsymbol{\theta}}^p_d + \boldsymbol{\Lambda} \boldsymbol{B} \boldsymbol{\tau} - \sum_{b \in \mathcal{B}} \boldsymbol{\Lambda} \boldsymbol{B} \boldsymbol{p}^a_{b,d}  -  \boldsymbol{\Lambda}_d \beta \boldsymbol{c}^p_d +  \boldsymbol{\Lambda}_d \beta \boldsymbol{t}^p_d + \boldsymbol{s}_d^a  \Bigr)    \notag\\
&+ \sum_{d \in \mathcal{D}} \sum_{b \in \mathcal{B}} \Bigl[ \overline{\boldsymbol{\phi}}_{b,d}^T \Bigl( \overline{\boldsymbol{e}}_{b,d} - \boldsymbol{p}^a_{b,d} \Bigr)  +  \underline{\boldsymbol{\phi}}_{b,d}^T \boldsymbol{p}^a_{b,d}  \Bigr]  + \sum_{d \in \mathcal{D}} \boldsymbol{\varphi}_{d}^T \boldsymbol{s}^a_{d}  \leq \iota, \label{const11_sle}
\end{align}
\end{subequations}
\noindent where the set of decision variables is $\Xi^{NRP} = \{ \boldsymbol{m}_{b,d}, \boldsymbol{p}_{b, d}^{a},  \boldsymbol{s}_d^a, \boldsymbol{\nu}_b, \boldsymbol{\rho}, \underline{\boldsymbol{\epsilon}}_d, \overline{\boldsymbol{\epsilon}}_d, \underline{\boldsymbol{\kappa}}_d, \overline{\boldsymbol{\kappa}}_d,$ $\underline{\boldsymbol{\phi}}_{b,d}, \overline{\boldsymbol{\phi}}_{b,d}, \boldsymbol{\varphi}_d, \beta, \boldsymbol{\tau} \}$. For the sake of simplicity, we assume blocks of identical length, i.e., $\overline{\boldsymbol{e}}_{b,d} = \left(\beta\overline{\boldsymbol{p}}^p_d + \boldsymbol{\tau} \right)/n_B$ for block $b$ and day $d$.

Expressions \eqref{fo_sle}--\eqref{const1bis_sle} represent the upper-level problem \eqref{fo_bilevel}--\eqref{const2_bilevel} while expressions \eqref{const2_sle}--\eqref{const11_sle} represent the regularized Karush–Kuhn–Tucker optimality conditions associated with the lower-level problems \eqref{forecasting_model}. 


In problem \eqref{single_level_equivalent}, we basically relax the sum of all complementary slackness conditions in \eqref{const11_sle} by means of the parameter $\iota > 0$. When this parameter is sufficiently small, we can speed up the search of a locally optimal solution found by a nonlinear solver. 

%

\subsection{Steps of the Proposed Approach}
\label{subsec:steps_proposed_approach}
A common shortcoming of any nonlinear program is its sensitivity to the initial search point. In order to avoid meaningless local optima, we propose the use of an efficient heuristic procedure in which two convex programs are sequentially run to infer the marginal utilities $\boldsymbol{m}_{b, d}$ and the homothetic parameters $\beta$ and $\boldsymbol{\tau}$. The result of this procedure can be utilized as a \textit{proxy} of the IO problem \eqref{bilevel_formulation}, and thus can be used to yield more interpretable local optimal solutions from the regularized nonlinear problem \eqref{single_level_equivalent}. The proposed heuristic procedure is built upon the one put forward in \cite{Saez-Gallego2018}, but has been substantially modified to account for the building thermal dynamics. For the sake of completeness, this procedure is fully described in \ref{sec:two_step_estimation}. 

Next, we list the steps of the proposed forecasting technique, which we denote as \textit{rnp}:

\begin{enumerate}
    \item We first solve the two-step heuristic estimation process described in \ref{sec:two_step_estimation} for a training set. This gives us a suitable \textit{proxy} for the solution of the original bilevel problem \eqref{bilevel_formulation}. From this procedure, we obtain the marginal utilities $\boldsymbol{m}_{b,d}$ and the corresponding estimates $\nu_b$ and $\boldsymbol{\rho}$, as well as the homothetic parameters $\widehat{\beta}$ and $\widehat{\boldsymbol{\tau}}$. 
    \item We then run the forecasting model \eqref{forecasting_model} for the training set to compute the value of the in-sample aggregate power $\boldsymbol{p}^a_{b,d}$, slack variable $\boldsymbol{s}^a_d$, and the dual variables.  
    \item Afterwards, we run the regularized nonlinear program \eqref{single_level_equivalent} with fixed homothetic parameters $\widehat{\beta}$ and $\widehat{\boldsymbol{\tau}}$. In addition, we take the solution from the forecasting model \eqref{forecasting_model} evaluated in the training set (previous step) as initialization. Here, we essentially \textit{re-optimize} the marginal utility curves with the aim of improving the solution given by the heuristics.
    \item Finally, the forecasting model \eqref{forecasting_model} is built with the homothetic parameters $\widehat{\beta}$ and $\widehat{\boldsymbol{\tau}}$ and the estimates $\widehat{\nu}_b$ and $\widehat{\boldsymbol{\rho}}$ for the marginal utility curves resulting from step 3 above.
\end{enumerate}

The parameters associated with the forecasting model \eqref{forecasting_model}, namely the marginal utilities as well as the homothetic parameters, can be periodically re-estimated with a new training data set in order to capture potential non-stationary patterns in the observed aggregate load.

\section{Comparison Methodologies}
\label{sec:comparison_methodologies}
We compare the forecasting capabilities of the proposed approach  against four benchmarks: (i) the nonlinear problem \eqref{single_level_equivalent} with $\beta$ and $\boldsymbol{\tau}$ free, without any initialization of the decision variables, and with $\iota = 0$, denoted as \textit{np w/o init}; (ii) a simpler two-step estimation procedure taken from \cite{Saez-Gallego2018} and denoted as \textit{s2s}; (iii) an AutoRegressive Integrated Moving Average Model with eXogenous (\textit{arimax}) variables; and (iv) a persistence model denoted as \textit{naive}. To the best of the authors' knowledge, \emph{arimax} models stand as the state-of-the-art black-box techniques to forecast the price-response of an aggregate of buildings \cite{corradi2012controlling}, as pointed out in Section \ref{sec:introduction}. 
The benefits of all models are compared by analyzing two error metrics on the aggregate power of the ensemble of buildings: the MAE and the root mean square error (RMSE) on a test data set. 

The forecasting problem associated with the two-step procedure \textit{s2s} is driven by the maximization of the welfare of the pool of buildings subject to solely the power bounds. In this benchmark, the indoor temperature bounds are ignored, thus overlooking the effect of the building thermal dynamics. Furthermore, in \textit{s2s},  the marginal utilities and the power bounds are inferred by successively running two linear programs so that the RMSE is minimized in a validation data set. The interested reader is referred to \cite{Saez-Gallego2018} for a detailed description of this methodology. 

The \textit{arimax} model has been implemented in Python  \cite{python} via the SARIMAX class of the package \textit{statsmodels}. We have set the maximum number of iterations to 1000 and the stopping rule is based on the estimator Akaike information criterion~(AIC). 

The forecast values of the aggregate power in day $d$ are equal to the observed values in the previous day $d-1$ for the  \textit{naive} model. The forecast error of this model is indicative of how hard predicting the demand of the pool of buildings is. 

\section{Case Study}
\label{sec:case_study}
We aim to learn the aggregate power of a population of 100 heterogeneous buildings. We first summarize the process to synthetically generate the data set. Then, we present the input data for testing the forecast capabilities. Subsequently, we discuss the results obtained with the proposed approach \textit{rnp} and the benchmarks. Finally, we also discuss the computational complexity of the proposed approach.

\subsection{Data Generation for a Pool of Buildings}
\label{subsec:data_generation_pool}
We assume that the consumption of each building $i$ for each day $d$ is driven by the following optimization problem:
\begin{subequations}
\label{data_generation_model}
\begin{align}
& \min_{p_{h}, s_{h}, \theta_{h}} \quad \sum_{h \in \mathcal{H}} \bigl( p_{h} \lambda_h + \varrho s_{h} \bigr) \label{obj_datagen}\\
& \theta_{h} = a_1 \theta_{h-1} + (1-a_1)\left[ \theta^{amb}_{h} - a_2 p_{h}\right], \quad  \forall h \in \mathcal{H} \label{const1_datagen}  \\
& - s_{h} + \underline{\theta}_{h} \leq \theta_{h} \leq \overline{\theta}_{h} + s_{h}, \quad \forall h \in \mathcal{H} \label{const2_datagen}  \\
& 0 \leq p_{h} \leq \overline{p}_{h}, \quad \forall h \in \mathcal{H} \label{const3_datagen} \\
& s_{h} \geq 0, \quad \forall h \in \mathcal{H}, \label{const4_datagen} 
\end{align}
\end{subequations}

\noindent where $\varrho$ represents a penalty cost related to the violation of the temperature-related constraints. Each building aims to minimize its electricity and penalty costs, as in \eqref{obj_datagen}, while satisfying the building thermal dynamics \eqref{const1_datagen}, the temperature comfort bounds \eqref{const2_datagen}, and the power bounds of the cooling device \eqref{const3_datagen}. Slack variables are declared non-negative in \eqref{const4_datagen}.

The technical parameters $\Omega^p$ for the prototype building are shown in Table~\ref{tab:data_prototype}. As done in \cite{zhao2017geometric}, the model parameters $\Omega^i$ for each building $i$ of the pool are assumed to be uniformly distributed based on a factor $\hbar$ modeling the degree of heterogeneity. For instance, the samples for the thermal capacitance $C^i$ are drawn from a uniform distribution in the interval $\bigl[ \bigl(1 -  \hbar\bigr) C^p , \bigl( 1 + \hbar \bigr) C^p \bigr]$, where $C^p$ is the thermal capacitance of the prototype building. The discretization step $\delta$ is assumed to be one hour and the penalty cost $\varrho$ is set to 0.01 \euro$/^\circ$C$\cdot$h for all buildings. Ambient temperature, electricity prices, and the building technical parameters $\Omega^i$ are given in \cite{Fernandez-Blanco2020}, for the sake of reproducibility. 

\begin{table}[h]
\caption{Technical Parameters $\Omega^p$ for the Prototype Building}
\label{tab:data_prototype}
\centering
\begin{tabular}{cccccccc}
\hline
$C$ (kWh$/^\circ$C)  & 10 && $\eta$ & 2.5 && $\Delta$ ($^\circ$C) & 1 \\
$R$ ($^\circ$C/kW) & 2 && $\theta^r$ ($^\circ$C) & 20 && $\theta_0$ ($^\circ$C) & 22.5 \\
$P$ (kW) & 5.4 && & && & \\
\hline
\end{tabular}
\end{table}

\subsection{Input Data for Testing the Forecasting Models}
\label{subsec:data}
We run simulations for 1872 hours (78 days) by using model \eqref{data_generation_model} for two different values of the heterogeneity factor: (i) $\hbar = 0.1$ (low heterogeneity among buildings), and (ii) $\hbar = 0.75$ (high heterogeneity among buildings). To avoid undesirable border effects, we disregard the results from the first day of the simulation. The sizes for the training, validation, and test sets are 35, 35, and 7 days, respectively, in chronological order. For each case, the aggregate power and the initial indoor air temperature per day can be found in \cite{Fernandez-Blanco2020}. Table \ref{tab:statistics_aggregate_power} summarizes some statistics on the aggregate power for both cases. The higher the degree of heterogeneity among buildings is, the smoother the power curve becomes. In other words, 10\% of heterogeneity leads to load synchronization with a maximum power peak of 541.0 kW and 61.8\% of periods where the buildings' load is 0. Conversely, 75\% of heterogeneity causes a lower peak, 218.4 kW, and a more distributed load. Assuming that the ambient temperature is perfectly forecast, we consider five regressors to estimate the marginal utility curves, namely the ambient temperature at hours $h-2$, $h-1$, $h$, $h+1$, and $h+2$, which are also reported in \cite{Fernandez-Blanco2020}. Finally, we consider that $\boldsymbol{c}^s$ is large enough, i.e., $\boldsymbol{c}^s=1$ for all time periods. 

\begin{table}[h]
\caption{Statistics on the Aggregate Power}
\label{tab:statistics_aggregate_power}
\centering
\begin{tabular}{ccc}
\cline{2-3} 
& $\hbar = 0.1$ & $\hbar = 0.75$ \\
\hline
Maximum (kW) & 541.0 & 218.4 \\
Mean (kW) & 64.0 & 42.0 \\
Total (MW) & 118.2 & 77.6 \\
\# hours without consumption (\%) & 61.8 & 0.0 \\
\hline
\end{tabular}
\end{table}

The simulations have been performed on a Windows-based computer with four CPUs clocking at 1.8 GHz and 8 GB of RAM. For the linear programs, we use CPLEX 12.8 \cite{Cplex} under Pyomo 3.7.3 \cite{Pyomo}. For the model $rnp$, we use the nonlinear solver CONOPT \cite{conopt} connecting to the NEOS server \cite{czyzyk_et_al_1998}.  

\subsection{Results}
\label{subsec:results}
We analyze the impact of the degree of heterogeneity of the pool of buildings on the forecasting capabilities of the proposed technique. Besides, for the models \textit{rnp}, \textit{np w/o init}, and \textit{s2s}, we further study the behavior of the models when considering either (i) a number of power blocks $n_B=1$, and (ii) $n_B=6$. Table \ref{tab:error_metrics} provides the forecast error metrics, namely RMSE and MAE, for all models outlined in Section~\ref{sec:comparison_methodologies} and the aforementioned cases. In this table, we highlight the best results in bold. 

\begin{table}[h]
\caption{Error Metrics -- Comparison of Models}
\label{tab:error_metrics}
\centering
\begin{tabular}{cc@{\hskip3pt}c@{\hskip6pt}c@{\hskip3pt}cc@{\hskip3pt}c@{\hskip3pt}c@{\hskip3pt}c}
\hline
\multirow{3}{*}{Model} & \multicolumn{4}{c}{$\hbar = 0.1$}  &  \multicolumn{4}{c}{$\hbar = 0.75$}  \\
\cline{2-9}
& \multicolumn{2}{c}{$n_B = 1$}  &  \multicolumn{2}{c}{$n_B = 6$} & \multicolumn{2}{c}{$n_B = 1$}  &  \multicolumn{2}{c}{$n_B = 6$} \\
\cline{2-9}
 & RMSE &  MAE&  RMSE &  MAE & RMSE &  MAE&  RMSE &  MAE \\
\hline
\textit{rnp}  & \textbf{106.7} & \textbf{52.7} & \textbf{103.7} & \textbf{52.5} &  \textbf{31.3} & \textbf{22.5} & \textbf{25.3} & \textbf{16.9}  \\
\textit{np w/o init}  & 165.3 & 76.3 & 165.3 & 76.3 & 35.9 & 22.9 & - & - \\
\textit{s2s}  & 132.8 & 87.5 & 133.0 & 88.9 &  38.4 & 24.0 & 30.3 &  26.0   \\
\textit{arimax}  & 161.8 & 108.3 & 161.8 & 108.3 & 31.6 & 23.0 & 31.6 & 23.0 \\
\textit{naive} & 177.5 & 90.4 & 177.5 & 90.4  & 36.9 & 24.2 & 36.9 & 24.2 \\
\hline
\end{tabular}
\end{table}

\begin{figure}[h]
\centerline{\includegraphics[scale=0.6]{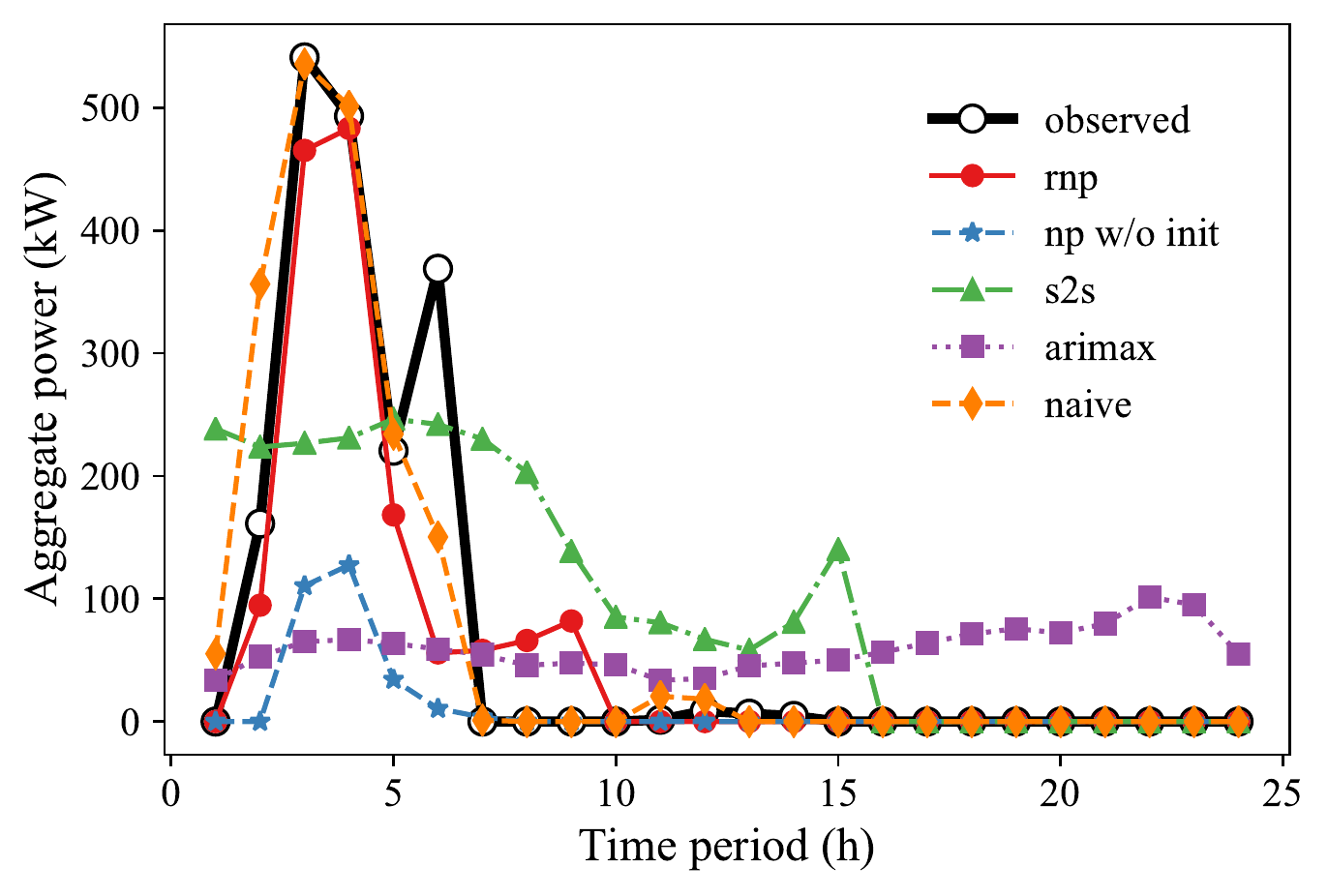}}
\vspace{-0.4cm}
\caption{Forecast and observed aggregate power for the first day of the test set with an heterogeneity factor of $0.1$ and $n_B = 1$.}
\label{fig:forecast_power_methods_het01_NB1}
\end{figure}

First, we discuss the results for an heterogeneity factor $\hbar = 0.1$ when considering a single power block, i.e. $n_B=1$. In this setup, the proposed method \textit{rnp} leads to the best forecasting performance in terms of RMSE and MAE since they are reduced by 39.9\% and 41.7\% compared to the \textit{naive} model. 
Solving the model \textit{np w/o init}, i.e. without fixing the homothetic parameters, without using any initialization, and with $\iota=0$, raises the RMSE and MAE by 54.9\% and 44.8\%, in that order, regarding the ones obtained with the proposed model \textit{rnp}. In Fig. \ref{fig:forecast_power_methods_het01_NB1}, we represent the forecast power for all benchmarks as well as the observed power of the first day of the test set. Unlike the method \textit{np w/o init}, we can observe that \textit{rnp} closely follows the observed power curve and is able to predict the peak periods thanks to the initialization.

The nonlinear model \textit{np w/o init} converges to the local optimal solution with $\beta = 0$, which seems to be an \textit{attractive} solution due to the nature of this mathematical problem. The reason behind this outcome relies on the definition of homothet \eqref{definition_homothet}. $\beta = 0$ implies a constant objective function in the lower-level problem \eqref{const3_bilevel} and a feasible region that boils down to the singleton $\boldsymbol{p}^a_d=\boldsymbol{\tau}, \forall d \in \mathcal{D}$, according to the definitions mentioned above. Therefore, the upper level \eqref{fo_bilevel}--\eqref{const2_bilevel} basically seeks the vector $\boldsymbol{\tau}$ minimizing the MAE over the training data set.

Both models \textit{s2s} and \textit{arimax} make substantially higher forecasting errors than the proposed technique \textit{rnp}, i.e. RMSE increases by 24.5\% and 51.6\%, in that order, whereas the respective increase in MAE is 66.0\% and 105.5\%. As seen in Fig. \ref{fig:forecast_power_methods_het01_NB1}, the method \textit{arimax} may work better for smoother processes and, as expected, it overlooks the irregularities of the aggregate power pattern. Method \textit{s2s} tends to adapt to the peak periods better than other benchmarks, although it is not successful in identifying them. The reason for those poor forecasts is that both models disregard the effect of the building thermal dynamics in the forecasting process. Finally, the \textit{naive} performs worse than any other technique in terms of RMSE for the test set, however its performance for the first day is not as bad as for the other days of the test set (see Fig.~\ref{fig:forecast_power_methods_het01_NB1}). 

For the case of $\hbar = 0.1$, when increasing the number of power blocks to 6, the proposed method \textit{rnp} refines the solution achieved with only one block, i.e., RMSE and MAE decreases by 2.8\% and 0.4\%, in that order. This is an indication that there is a small degree of sensitiveness of the power to the price, which is captured by means of the marginal utilities in the proposed forecasting model. 

\begin{figure}[h]
\centerline{\includegraphics[scale=0.6]{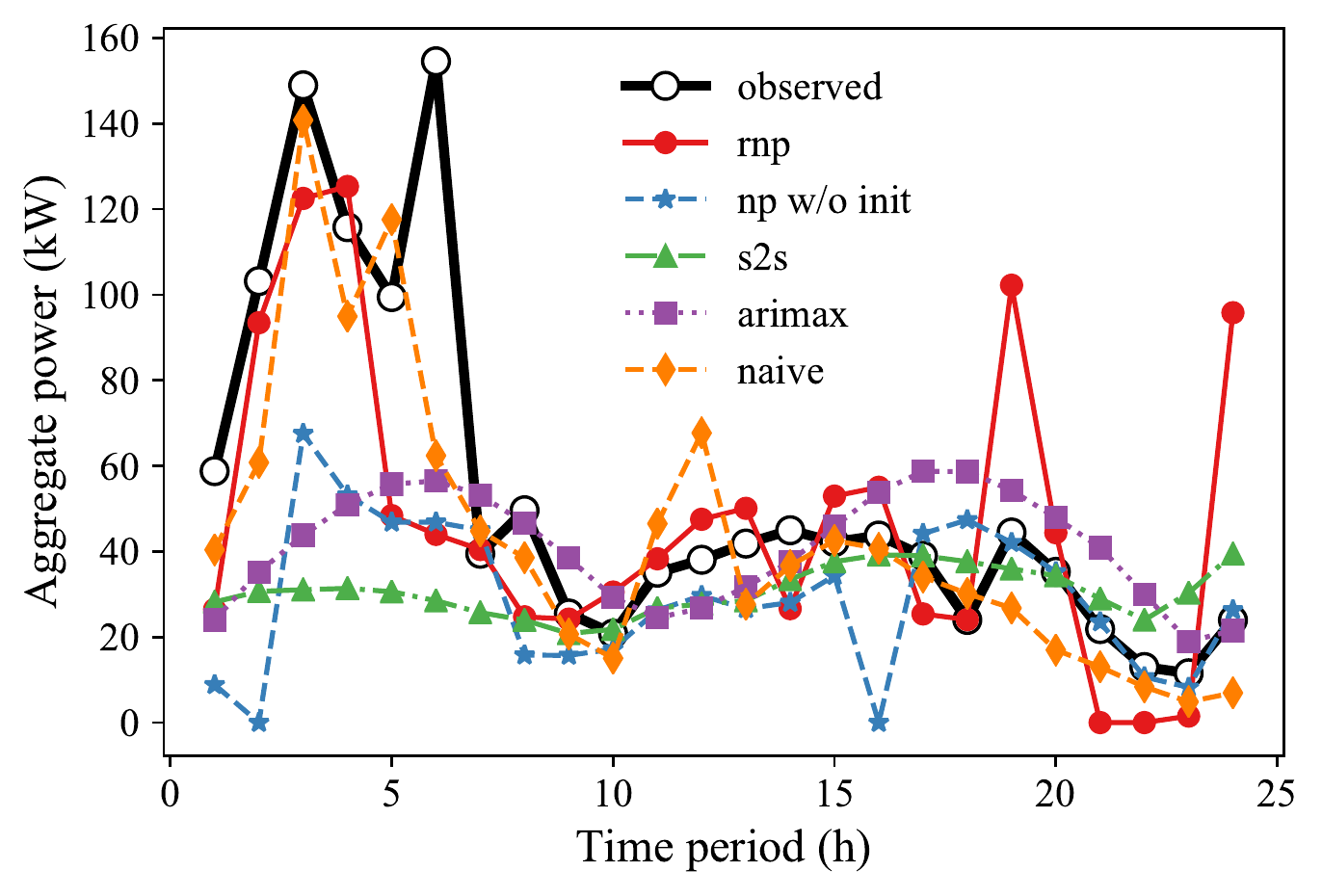}}
\vspace{-0.4cm}
\caption{Forecast and observed aggregate power for the first day of the test set with an heterogeneity factor of $0.75$ and $n_B = 1$.}
\label{fig:forecast_power_methods_het075_NB1}
\end{figure}

Fig. \ref{fig:forecast_power_methods_het075_NB1} illustrates the forecast and observed aggregate power for all methods for the first day of the test set with $\hbar=0.75$ and $n_B = 1$. We can see that now the power forecasts of all benchmarks are more alike because the aggregate power becomes smoother along the time for the high-heterogeneity case. Therefore, we need to resort to Table \ref{tab:error_metrics} so we can quantify the forecast error in terms of the error metrics in the test set. 
Due to the smoothness of the aggregate power, the \textit{naive} model provides a better accuracy compared to the results with a low heterogeneity factor, i.e. RMSE = 36.9 and MAE = 24.2. In terms of RMSE, the proposed technique \textit{rnp} exhibits the best forecasting performance with $n_B = 1$, which results in a 15.2\%, 18.5\%, 12.8\%, and 0.9\% of improvement over the error metrics attained by the \textit{naive}, \textit{s2s}, \textit{np w/o init}, and \textit{arimax} models, in that order. However, this improvement over the \textit{naive} model (15.2\% in RMSE) is substantially lower than when the buildings are more alike, wherein there is a reduction of 39.9\% in RMSE.

\begin{figure}[h]
\centerline{\includegraphics[scale=0.6]{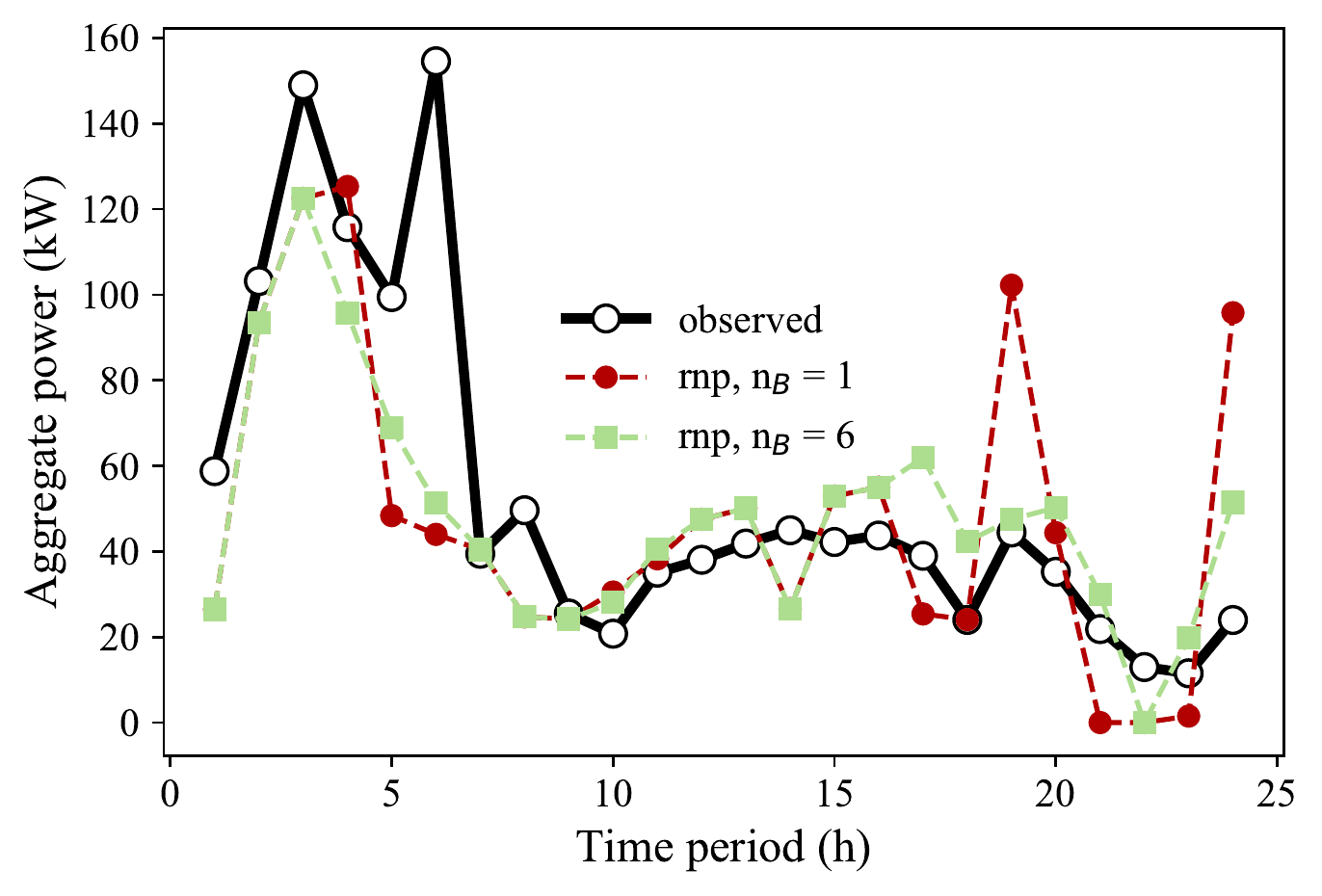}}
\vspace{-0.4cm}
\caption{Forecast and observed aggregate power for the model \textit{rnp} for an heterogeneity factor of $0.75$ with both $n_B = 1$ and $n_B = 6$.}
\label{fig:forecast_power_methods_het075_model_rnp}
\end{figure}

In addition, for the case of high heterogeneity, the forecasting performance of the proposed technique considerably enhances the error metrics when increasing the number of blocks to $n_B=6$. Specifically, there is a reduction of 19.2\% in RMSE and 24.9\% in MAE over the solution with only one block, whereas this reduction is just 2.8\% and 0.4\%, respectively, for $\hbar = 0.1$. We show, for the first day of the test set, the power forecast provided by \textit{rnp} in the high-heterogeneity case when considering either $n_B=1$ or $n_B=6$ in Fig. \ref{fig:forecast_power_methods_het075_model_rnp}. Some forecast values are improved due to the power-price sensitiveness captured by the increasing number of blocks and, therefore, they are close to their respective observed values. 

A similar observation can be made for \textit{s2s}, which also captures the price-responsiveness of the pool of buildings by estimating a step-wise utility function. However, the forecasting capabilities of the model \textit{s2s}, which is the one proposed in \cite{Saez-Gallego2018}, are by far worse than those from the proposed technique due to the fact that the former ignores the building thermal dynamics in the forecasting process. Finally, the model \textit{np w/o init} is unable to find a solution and the solver CONOPT returns an evaluation error. This is probably caused because model~\eqref{single_level_equivalent} with $\iota=0$ is inherently ill-posed, as pointed out in \cite{scholtes2001convergence}.

\subsection{Computational Complexity}
\label{subsec:comp_complexity}
Table \ref{tab:comp_complexity} presents the computing times for solving the estimation problem, i.e. the regularized nonlinear single-level program \eqref{single_level_equivalent} for the proposed model \emph{rnp} and the model \textit{np w/o init}, for all cases analyzed. The estimates for the proposed model \emph{rnp}, i.e. when decision variables are initialized, are achieved in less than 20 min, which is acceptable for a day-ahead operational problem. We can also observe that the model \textit{np w/o init} without initialization attains a solution faster, albeit worse in terms of prediction error, than the proposed model. This is due to the convergence of \textit{np w/o init} to the \emph{attractive} local optimal solution with $\beta = 0$, as explained in the previous section. Moreover, the computational burden of the proposed \emph{rnp} is strongly related to the number of blocks of the marginal utility, which plays a major role for high heterogeneity factors. Note that the computing times for solving the forecasting linear problem \eqref{forecasting_model} are negligible.

On the other hand, the computational performance of this forecasting approach is scalable for real-sized aggregators with thousands of buildings for two reasons. First, models \eqref{forecasting_model} and \eqref{single_level_equivalent} are independent of the number of buildings belonging to the aggregator. Second, both models are built around a single \emph{prototype} building, whose feasible region is geometrically transformed via the homothetic parameters, also independent of the size of the pool.

\begin{table}[h]
\caption{Computing Times (s) for the Estimation Problem \eqref{single_level_equivalent} for Models \textit{rnp} and \textit{np w/o init}}
\label{tab:comp_complexity}
\centering
\begin{tabular}{cc@{\hskip3pt}c@{\hskip6pt}c@{\hskip3pt}cc@{\hskip3pt}c@{\hskip3pt}c@{\hskip3pt}c}
\hline
\multirow{2}{*}{Model} & \multicolumn{2}{c}{$\hbar = 0.1$}  &  \multicolumn{2}{c}{$\hbar = 0.75$}  \\
\cline{2-5}
& \multicolumn{1}{c}{$n_B = 1$}  &  \multicolumn{1}{c}{$n_B = 6$} & \multicolumn{1}{c}{$n_B = 1$}  &  \multicolumn{1}{c}{$n_B = 6$} \\
\hline
\textit{rnp}  &  46.3 & 933.7  & 43.6  & 555.4   \\
\textit{np w/o init}  & 41.9 & 153.3  & 30.8  & - \\
\hline
\end{tabular}
\end{table}

\section{Conclusion}
\label{sec:conclusion}
This paper has proposed a novel day-ahead forecasting technique for an aggregation of smart buildings equipped with thermostatically-controlled loads. From a modeling perspective, the aggregate power of the pool of buildings is represented by using a geometric approach, i.e., its price-response is characterized by a set of marginal utility curves and a homothet of a prototype building. This intuitive representation of the aggregate allows us to account for the building thermal dynamics while drastically reducing the number of parameters to be estimated. Hence, the computational complexity of the estimation algorithm is decreased, thus avoiding the undesirable overfitting effect. From a methodological perspective, inverse optimization is applied to infer the marginal utilities and the homothetic parameters by means of bilevel programming. We can conclude that (i) accounting for the building thermal dynamics in the forecasting technique improves the forecasting error by 20--40\% compared to existing and persistence methodologies when the buildings are more alike, and that (ii) the use of an increasing number of blocks for the marginal utilities in the forecasting process substantially improves the accuracy of the proposed forecasting technique when the heterogeneity among buildings is high. 

This paper prompts several avenues for future research. Further work will be devoted to improving the forecast accuracy. Three potential extensions of this research are: (i) increasing the number of geometric parameters (e.g. a rotation of the homothet), (ii) extending the geometric parameters to be regressor-dependent, and (iii) combining inverse optimization with machine learning techniques. Besides, we will explore the impact of inverse optimization under noisy data. Finally, intra-day forecasting of an aggregate load of an ensemble of buildings is also an interesting direction of future research.

\appendix
\section{Heuristic Estimation Approach}
\label{sec:two_step_estimation}
The proposed two-step heuristic estimation method is built upon the one proposed in \cite{Saez-Gallego2018}. However, the procedure has been substantially modified to account for the building thermal dynamics. This estimation process is made up of two linear problems: (i) the \textit{feasibility problem} and (ii) the \textit{optimality problem}. The respective formulations are described in \ref{sec:feasibility_problem} and \ref{sec:optimality_problem}. Then, we list the steps of the heuristic estimation approach in \ref{sec:steps}.

\subsection{Feasibility Problem}
\label{sec:feasibility_problem}
The feasibility problem can be formulated as: 
\begin{subequations}
\label{feasibility_problem}
\begin{align}
&\min_{\Xi^{FP}} \quad \sum_{d \in \mathcal{D}} \left( \overline{\boldsymbol{\xi}}^{+}_d + \underline{\boldsymbol{\xi}}^{+}_d + \overline{\boldsymbol{\chi}}^{+}_d + \underline{\boldsymbol{\chi}}^{+}_d \right) \left(1 - H \right) + \sum_{d \in \mathcal{D}} \left( \overline{\boldsymbol{\xi}}^{-}_d + \underline{\boldsymbol{\xi}}^{-}_d + \overline{\boldsymbol{\chi}}^{-}_d + \underline{\boldsymbol{\chi}}^{-}_d \right) H \label{fp_constr1}\\
& \boldsymbol{p}^{a^\prime}_{d} - \beta \underline{\boldsymbol{p}}^p_d - \boldsymbol{\tau}  = \underline{\boldsymbol{\xi}}^{+}_d - \underline{\boldsymbol{\xi}}^{-}_d, \quad \forall d \in \mathcal{D} \label{fp_constr2}\\
& \beta\overline{\boldsymbol{p}}^p_d + \boldsymbol{\tau} - \boldsymbol{p}^{a^\prime}_{d} = \overline{\boldsymbol{\xi}}^{+}_d - \overline{\boldsymbol{\xi}}^{-}_d, \quad  \forall d \in \mathcal{D} \label{fp_constr3}\\
& \boldsymbol{\Lambda} \boldsymbol{B} \boldsymbol{p}^{a^\prime}_d + \boldsymbol{\Lambda} \beta \left(\boldsymbol{c}^p_d \hspace{-0.1cm} + \hspace{-0.1cm} \boldsymbol{t}^p_d\right) - \beta \underline{\boldsymbol{\theta}}^p_d  - \boldsymbol{\Lambda} \boldsymbol{B}  \boldsymbol{\tau} =  \underline{\boldsymbol{\chi}}^{+}_d - \underline{\boldsymbol{\chi}}^{-}_d, \quad  \forall d \hspace{-1pt}  \in \hspace{-1pt}  \mathcal{D} \label{fp_constr4}\\
& \beta \overline{\boldsymbol{\theta}}^p_d + \boldsymbol{\Lambda} \boldsymbol{B} \boldsymbol{\tau} -  \boldsymbol{\Lambda} \boldsymbol{B} \boldsymbol{p}^{a^\prime}_d - \boldsymbol{\Lambda} \beta \left(\boldsymbol{c}^p_d \hspace{-0.1cm} + \hspace{-0.1cm} \boldsymbol{t}^p_d\right)  = \overline{\boldsymbol{\chi}}^{+}_d - \overline{\boldsymbol{\chi}}^{-}_d, \quad  \forall d  \in \mathcal{D} \label{fp_constr5} \\
& \overline{\boldsymbol{\xi}}^{+}_d,  \underline{\boldsymbol{\xi}}^{+}_d,  \overline{\boldsymbol{\chi}}^{+}_d, \underline{\boldsymbol{\chi}}^{+}_d,  \overline{\boldsymbol{\xi}}^{-}_d,  \underline{\boldsymbol{\xi}}^{-}_d, \overline{\boldsymbol{\chi}}^{-}_d, \underline{\boldsymbol{\chi}}^{-}_d \geq 0,\quad   \forall d \in \mathcal{D}, \label{fp_constr7}
\end{align}
\end{subequations}

\noindent where the set of decision variables $\Xi^{FP} = \{ \overline{\boldsymbol{\xi}}^{+}_d,  \underline{\boldsymbol{\xi}}^{+}_d,  \overline{\boldsymbol{\chi}}^{+}_d, \underline{\boldsymbol{\chi}}^{+}_d,  \overline{\boldsymbol{\xi}}^{-}_d,  \underline{\boldsymbol{\xi}}^{-}_d,\overline{\boldsymbol{\chi}}^{-}_d,\underline{\boldsymbol{\chi}}^{-}_d, \beta, \boldsymbol{\tau} \}$. 

The objective function \eqref{fp_constr1} minimizes the sum of feasibility and infeasibility slack variables associated with the feasible region of forecasting problem \eqref{forecasting_model}. Constraints \eqref{fp_constr2}--\eqref{fp_constr3} represent the power bound limits with the feasibility and infeasibility slack variables. Likewise, expressions \eqref{fp_constr4}--\eqref{fp_constr5} model the temperature bounds including the aforementioned slack variables. In \eqref{fp_constr2}--\eqref{fp_constr5}, $\boldsymbol{p}^{a^\prime}_{d}$ is the vector of observed aggregate power values for day $d$. Constraints \eqref{fp_constr7} impose the non-negativity of slack variables. Finally, it is worth emphasizing that $H$ could be a control parameter ranging in the interval $[0, 1)$. However, we assume that $H$ is large enough, i.e. $H=0.99$, to allocate as much observed power as possible within the bounds of the feasible region of the forecasting problem \eqref{forecasting_model}.

\subsection{Optimality Problem}
\label{sec:optimality_problem}
The optimality problem can be formulated as: 
\begin{subequations}
\label{optimality_problem}
\begin{align}
&\min_{\Xi^{OP}} \quad \sum_{d \in \mathcal{D}} \lvert \varepsilon_d \rvert \label{op_constr1} \\
& \text{subject to:} \notag\\
& \boldsymbol{m}_{b, d} = \boldsymbol{\nu}_b + \boldsymbol{Z}_{d}\boldsymbol{\rho} , \quad \forall b \in \mathcal{B}, d \in \mathcal{D} \label{op_constr2}  \\
& \boldsymbol{\nu}_b \geq \boldsymbol{\nu}_{b+1}, \quad \forall b < n_B \label{op_constr3}\\
& - \underline{\boldsymbol{\epsilon}}_d^T \bigl( \widehat{\beta} \underline{\boldsymbol{p}}^p_d + \widehat{\boldsymbol{\tau}} \bigr)  + \overline{\boldsymbol{\epsilon}}_d^T \bigl( \widehat{\beta}\overline{\boldsymbol{p}}^p_d + \widehat{\boldsymbol{\tau}} \bigr)   - \underline{\boldsymbol{\kappa}}_d^T \left(\widehat{\beta} \underline{\boldsymbol{\theta}}^p_d  + \boldsymbol{\Lambda} \boldsymbol{B}  \widehat{\boldsymbol{\tau}} -  \boldsymbol{\Lambda} \widehat{\beta} \left(\boldsymbol{c}^p_d \hspace{-0.1cm} + \hspace{-0.1cm} \boldsymbol{t}^p_d\right) \right) \notag \\ 
& + \overline{\boldsymbol{\kappa}}_d^T \left( \widehat{\beta} \overline{\boldsymbol{\theta}}^p_d + \boldsymbol{\Lambda} \boldsymbol{B} \widehat{\boldsymbol{\tau}} - \boldsymbol{\Lambda} \widehat{\beta} \left(\boldsymbol{c}^p_d \hspace{-0.1cm} + \hspace{-0.1cm} \boldsymbol{t}^p_d\right)  \right) + \sum_{b \in \mathcal{B}}  \overline{\boldsymbol{\phi}}_{b,d}^T \widehat{\overline{\boldsymbol{e}}}_{b,d}   - \varepsilon_d \notag \\ 
&  = \sum_{b \in \mathcal{B}}\bigl(\boldsymbol{m}_{b,d}^T - \boldsymbol{\lambda}^T_d \bigr) \boldsymbol{p}^{a^{\prime}}_{b,d}  - \boldsymbol{c}^{s, T}_d \boldsymbol{s}^{a}_d,  \quad \forall d \in \mathcal{D}  \label{op_constr4} \\
& - \underline{\boldsymbol{\epsilon}}_d + \overline{\boldsymbol{\epsilon}}_d - \boldsymbol{B}^T \boldsymbol{\Lambda}^T \underline{\boldsymbol{\kappa}}_d + \boldsymbol{B}^T \boldsymbol{\Lambda}^T \overline{\boldsymbol{\kappa}}_d \notag \\ 
& + \overline{\boldsymbol{\phi}}_{b,d} - \underline{\boldsymbol{\phi}}_{b,d} =   \boldsymbol{m}_{b,d} - \boldsymbol{\lambda}_d, \quad \forall d \in \mathcal{D}, b \in \mathcal{B} \label{op_constr5}  \\
& \underline{\boldsymbol{\kappa}}_d +  \overline{\boldsymbol{\kappa}}_d  \leq  \boldsymbol{c}^{s}_d, \quad \forall b \in \mathcal{B}, d \in \mathcal{D} \label{op_constr6}  \\
&  \widehat{\beta} \underline{\boldsymbol{\theta}}^p_d  + \boldsymbol{\Lambda} \boldsymbol{B}  \widehat{\boldsymbol{\tau}}  -  \boldsymbol{s}^a_d  \leq  \hspace{-0.1cm} \sum_{b \in \mathcal{B}} \hspace{-0.1cm} \boldsymbol{\Lambda} \boldsymbol{B} \boldsymbol{p}^{a^{\prime}}_{b, d} \hspace{-0.1cm} + \hspace{-0.1cm} \boldsymbol{\Lambda} \widehat{\beta} \left(\boldsymbol{c}^p_d \hspace{-0.1cm} + \hspace{-0.1cm} \boldsymbol{t}^p_d\right) \hspace{-0.1cm} \leq \widehat{\beta}\, \overline{\boldsymbol{\theta}}^p_d + \boldsymbol{\Lambda} \boldsymbol{B} \widehat{\boldsymbol{\tau}} \hspace{-0.1cm}+ \hspace{-0.1cm} \boldsymbol{s}^a_d,  \forall d \in \mathcal{D}  \label{op_constr7}  \\
& \boldsymbol{s}^a_d, \underline{\boldsymbol{\epsilon}}_d, \overline{\boldsymbol{\epsilon}}_d, \underline{\boldsymbol{\kappa}}_d, \overline{\boldsymbol{\kappa}}_d \geq 0, \quad \forall d \in \mathcal{D} \label{op_constr8}\\
& \underline{\boldsymbol{\phi}}_{b, d}, \overline{\boldsymbol{\phi}}_{b, d} \geq 0, \quad \forall b \in \mathcal{B}, \forall d \in \mathcal{D}, \label{op_constr9}
\end{align}
\end{subequations}

\noindent where the set of decision variables $\Xi^{OP} = \{ \varepsilon_d, \boldsymbol{m}_{b,d}, \nu_b, \boldsymbol{\rho}, \underline{\boldsymbol{\epsilon}}_d, \overline{\boldsymbol{\epsilon}}_d, \underline{\boldsymbol{\kappa}}_d, \overline{\boldsymbol{\kappa}}_d, \underline{\boldsymbol{\phi}}_{b, d}, \overline{\boldsymbol{\phi}}_{b, d}\}$. 

The objective function \eqref{op_constr1} minimizes the absolute value of the sum of the duality gaps per day of problem \eqref{forecasting_model}. As similarly done in the bilevel problem \eqref{bilevel_formulation}, marginal utilities are modeled as linear regression functions in terms of explanatory variables via constraints \eqref{op_constr2}, whereas constraints \eqref{op_constr3} characterize them to be monotonically non-increasing. We assume that vector $\boldsymbol{\nu}_b$ is time invariant, i.e. all components $\nu_{b, h}$ are identical. Constraints \eqref{op_constr4} are the relaxed equality constraints associated with the strong duality theorem. Constraints \eqref{op_constr5}--\eqref{op_constr6} are the dual feasibility constraints. We must also add the primal feasibility constraints \eqref{op_constr7} related to the temperature bounds since the slack variables become now decision variables. Finally, constraints \eqref{op_constr8}--\eqref{op_constr9} declare the non-negativity of primal and dual decision variables.

\subsection{Steps for the Heuristic Approach}
\label{sec:steps}
After presenting the feasibility and optimality problems in the previous sections, we list the steps for the heuristic estimation approach: 
\begin{enumerate}
    \item We first solve the \textit{feasibility problem}. Its main goal is to estimate the parameters shaping the feasible region of the forward problem \eqref{forecasting_model}, by using the observed aggregate values of the pool of buildings. In this paper, the feasible region exclusively depends on the homothetic parameters $\beta$ and $\boldsymbol{\tau}$. The formulation of the feasibility problem can be found in the previous \ref{sec:feasibility_problem}.  
    \item Once the homothetic parameters $\widehat{\beta}$ and $\widehat{\boldsymbol{\tau}}$ are inferred from the feasibility problem, we can compute the power block upper limit $\widehat{\overline{\boldsymbol{e}}}_{b,d}$, $\forall b \in \mathcal{B}$, according to the assignments shown in Table \ref{tab:power_block_limits} wherein $\widehat{\underline{p}}^a_h = \widehat{\beta} \underline{p}^p_h + \widehat{\tau}_h$ and $\widehat{\overline{p}}^a_h = \widehat{\beta} \overline{p}^p_h + \widehat{\tau}_h$. For instance, let us assume three power blocks, i.e., $n_B=3$, and that the estimated minimum and maximum aggregate power bounds are $\widehat{\underline{p}}^a = 10$ kW and  $\widehat{\overline{p}}^a=40$ kW. For the sake of simplicity, we have dropped indices $h$ and $d$. Given the assignments in Table \ref{tab:power_block_limits} (last column and rows 3 and 4), we can set $\widehat{\overline{e}}_1=\widehat{\underline{p}}^a = 10$ kW for the first block, and $\widehat{\overline{e}}_2=\widehat{\overline{e}}_3=\dfrac{\widehat{\overline{p}}^a-\widehat{\underline{p}}^a}{n_B-1} = 15$ kW for the remaining blocks. 
    \item Then, the estimated homothetic parameters $\widehat{\beta}$ and $\widehat{\boldsymbol{\tau}}$ can be plugged into the  \textit{optimality problem}, which can be derived by using results from duality theory of linear programming \cite{luenberger1984linear}. This optimality problem aims to estimate the parameters of forward problem \eqref{forecasting_model} that are related to the optimality of the observed power values, i.e., the marginal utilities $\boldsymbol{m}_{b,d}$. The formulation of the optimality problem can be found in the previous \ref{sec:optimality_problem}.
\end{enumerate}

\setcounter{table}{0} \renewcommand{\thetable}{A.\arabic{table}} 
\begin{table}[t]
\caption{Value of Power Block Upper Limit $\widehat{\overline{e}}_{h,b}$}
\label{tab:power_block_limits}
\centering
\begin{tabular}{c@{\hspace{1\tabcolsep}}c@{\hspace{1\tabcolsep}}c@{\hspace{1\tabcolsep}}c@{\hspace{1\tabcolsep}}c}
\cline{3-5}
\\[-14pt]
\multicolumn{2}{c}{} &  $\widehat{\underline{p}}^a_h \leq 0 \leq \widehat{\overline{p}}^a_h$ &  $\widehat{\overline{p}}^a_h \leq 0$  &  $\widehat{\underline{p}}^a_h \geq 0$ \\[3pt]
\hline 
\multirow{1}{*}{$n_B=1$} & $b \in \mathcal{B}$  & $\widehat{\overline{p}}^a_h$  & $0$ & $\widehat{\overline{p}}^a_h$ \\[3pt]
\hline
\multirow{2}{*}{$n_B > 1$} & $b=1$  & $\widehat{\overline{p}}^a_h/n_B$ & $0$ & $\widehat{\underline{p}}^a_h$ \\[3pt]
                                       & $b \in \mathcal{B} \setminus \{1\}$  & $\widehat{\overline{p}}^a_h/n_B$  & $0$  & $\frac{\left(\widehat{\overline{p}}_h^a - \widehat{\underline{p}}_h^a\right)}{n_B - 1}$ \\
\hline
\end{tabular}
\end{table}

\section*{Acknowledgements}
This project has received funding in part by the Spanish Ministry of Economy, Industry, and Competitiveness through project ENE2017-83775-P; and in part by the European Research Council (ERC) under the European Union's Horizon 2020 research and innovation programme (grant agreement No 755705).


\bibliography{homothetic}

\end{document}